\documentclass[12pt]{article}

\usepackage[cp1251]{inputenc}
\usepackage[centertags]{amsmath}
\usepackage{amsfonts}
\usepackage{amssymb}
\usepackage{amsthm}
\usepackage{amsmath}
\usepackage{newlfont}
\usepackage{graphicx}

\date{}
\title{Spectroscopy of new c,b-mesons}
\author{S. M. Gerasyuta$^{1,2}$, M. A. Durnev$^{1}$ \\  {\em  $^{1}$Department of Theoretical Physics }\\ 
\em {St. Petersburg State University, 198904,} \\
\em {St. Petersburg, Russia}\\
{\em $^{2}$Department of Physics, LTA, 194021, St. Petersburg, Russia}
} 
\scrollmode

\begin{document}
\maketitle

\begin{abstract}
{\small 
\noindent Masses of the lowest multiplets of $c,b$-mesons in $S$- and $P$-wave states with quantum numbers
$J^{PC}: 0^{-+}, 1^{--}, 0^{++}, 1^{+-}, 1^{++}, 2^{++}$ are obtained with the help of dispersion $N/D$
method of heavy quark effective theory. The results are in a good agreement with experimental data. 
Radiative decay widths for some states mentioned above were calculated.    
}
\end{abstract}
{\bf 1. Introduction \\}

The mass spectrum of the charmed and beauty mesons is the subject of investigation of many
theoretical papers using the different approaches. The unquestionable advantage of the potential
approach to the description of heavy mesons is its simplicity and clarity. One can
calculate the masses, the radiative  and the annihilation decay widths using
a chosen potential [1-7]. This allows to determine the meson wave functions in a wide
range of distances and to describe the quark interaction dynamics. The method of QCD sum rules
has been successfully used to determine the masses of the light and heavy mesons [8,9]. Owing
to a small relative difference of the resonance masses the bottomonium contribution of 
$\Upsilon$-meson and ground states in other channels dominates only for the momenta with $n \ge 20$. 
The most difficulty of applying the sum rule method 
for bottomonium - is to take into account the relativistic corrections,  i.e. the 
next term in the expansion in $n^{-1}$.  For this it is necessary to sum up all the terms 
of the type of $(\alpha_{s}n^{1/2})^{k}n^{-1}$, or $\alpha_{s}^{2}(\alpha_{s}n^{1/2})^{k}$ 
(since $\alpha_{s}n^{1/2} \sim 1$), which side by side with the pure relativistic effects 
(described by the Breit-Fermi Hamiltonian) contain the radiation corrections of the order 
of $\alpha_{s}^{2}$. This difficulty reveals, for example, in the calculating of masses  
$1P$-bottomonium levels. In this case the quark model can be used, which takes into 
account the relativistic effects of the quark interaction dynamics. 

After the discovery of the $D_{sJ}^{*}(2317)$ [10] and $D_{sJ}(2460)$ [11] mesons a number of 
experimental investigations of these narrow resonances was carried out [10-14].  Their 
masses are 100 MeV smaller than those predicted by the quark model. A large number of theoretical 
papers was devoted to the study of structure of these mesons. The interpretation of the $D_{sJ}^{*}(2317)$, 
$D_{sJ}(2460)$ as the lowest $P$-wave ${\bar c}s$-states [15, 16] is natural. However another 
nature of these states is also supposed: $DK$-molecule [17],  tetraquark states [18, 19],  
$D\pi$-atom [20] and others. In our study these states are considered as the ${\bar c}s$-mesons.
The masses of all states with the open charm and the charmonium states  with the quantum  
numbers $J^{PC}:0^{-+},  1^{--}, 0^{++}, 1^{+-}, 1^{++}, 2^{++}$ have been calculated (Table 1).  
The analogous computations have been carried out to the $b$-mesons including the $u,d,s,c,b$-quarks 
(Table 2). We have used the dispersion N/D method of heavy quark effective theory.

Section II is devoted to the computation of the mass spectrum of $c,b$-mesons with the quantum
numbers $J^{PC}=0^{-+}, 1^{--}, 0^{++}, 1^{+-}, 1^{++}, 2^{++}$ with the help of the
dispersion N/D method of heavy quark effective theory (Tables 1, 2).

The radiative decay widths of heavy mesons are calculated in Section III with the help of the
dispersion integration method.

The results of the $c,b$-meson characteristics computation and the problems, which remain
open, are discussed in the Conclusion. 
\\ \\ \\

{\bf 2. N/D method in heavy quark effective theory}\\

The amplitude of the $N/D$ method is defined as:
{\multlinegap=0pt \begin{multline}
\qquad \qquad  \qquad \qquad A(s)=\dfrac {N(s)} {D(s)}, \quad D(s)=1-B(s), 
\qquad \qquad \qquad   \tag{1} \end{multline} }
\\[-30pt]
{\multlinegap=0pt \begin{multline}
\qquad \qquad  \qquad \qquad B(s)= \int\limits_{(m_{1}+m_{2})^2} ^{\Lambda} {ds' 
\dfrac {\rho(s')G(s')} {s' - s}}.
\qquad \qquad \qquad \qquad  \qquad  \tag{2} \end{multline} }

Here  $m_{1}$ is the heavy quark $Q (Q=c,b)$ mass, $m_{2}$ is the light quark $q$ \; $(q=u,d,s; \; q\bar Q-$meson) 
mass or the heavy quark $Q \; (Q=c,b; \; Q\bar Q-$meson; $m_{2} \le m_{1}$) mass, 
$\Lambda$ is the cutoff parameter, $\rho(s)$ is the two-particle phase space, which is given
by formula [21-23]:
{\multlinegap=0pt \begin{multline}
 \rho(s)= \Bigl(\alpha \dfrac {s} {(m_{1}+m_{2})^2}+\beta+\frac {1} {s}
\delta \Bigr)  \dfrac{\sqrt{(s-(m_{1}+m_{2})^2)(s-(m_{1}-m_{2})^2)}}{s}, \quad  
 \tag{3} \end{multline} }
$\alpha, \beta, \delta$ are the coefficients, which are different for different states (Table 3),
$G(s)$ are smooth functions of $s$, which depend on the quantum numbers $J^{PC}$ of the states (Table 4).
The amplitude $A(s)$ has only one singularity point $s=(m_{1}+m_{2})^2$, therefore using the small
value of ratio $\dfrac {m_{2}} {m_{1}}$ ($m_{2}$ is the light quark mass) Eq.(3) can be rewritten
in the form:
{\multlinegap=0pt \begin{multline}
\qquad \qquad \rho(s)= \left.\left(\alpha \dfrac {s} {(m_{1}+m_{2})^2}+\beta+\dfrac {1} {s}
\delta \right) \right. \sqrt{\dfrac {s-(m_{1}+m_{2})^2} {s}}. \quad \qquad    
\tag{4} \end{multline} }
We take into account, that the expression $\sqrt{\dfrac {s-(m_{1}-m_{2})^2} {s}} \sim 1$ in the
middle point of the physical region $(m_{1}+m_{2})^2 \div \Lambda$. We obtain an expression
for the function $B(s)$, which describes accurately the heavy meson spectrum and has also the same
structure as the explicit expression for $B$-function [21-23]:  \\[-10pt]
{\multlinegap=0pt \begin{multline}  
B(s) = G(s) \Biggl[   \left.\left (\alpha \dfrac {s} {(m_{1}+m_{2})^2}+\beta+\dfrac {1} {s}
\delta \right) \right. \sqrt{\dfrac {s-(m_{1}+m_{2})^2}{s}}\times \qquad \qquad
  \\[10pt]
\shoveleft{\times 
\Biggl(\ln {\dfrac {\sqrt{\frac{\Lambda-
(m_{1}+m_{2})^2} {\Lambda}}-\sqrt{\frac{s-(m_{1}+m_{2})^2} {s}}} {
\sqrt{\frac{\Lambda-(m_{1}+m_{2})^2} {\Lambda}}+\sqrt{\frac{s-(m_{1}+m_{2})^2} {s}}}} +i\pi \Biggr)  
 + \alpha \dfrac {\sqrt{\Lambda(\Lambda-(m_{1}+m_{2})^2)}} {(m_{1}+m_{2})^2}  + } \\[10pt]
 \shoveleft{+ 2 \dfrac
{\delta} {s} \sqrt{\dfrac {\Lambda-(m_{1}+m_{2})^2}{\Lambda}}  + \left.
\left(\beta+\alpha \bigl (\dfrac {s} {(m_{1}+m_{2})^2}-\dfrac {1} {2} \bigr ) \right) \right.
\ln {\dfrac {1+\sqrt{\frac {\Lambda-(m_{1}+m_{2})^2}{\Lambda} }} {1-\sqrt{\frac {
\Lambda-(m_{1}+m_{2})^2} {\Lambda} }}} \ \Biggr] } 
\raisetag{95pt} \tag{5} \end{multline} }

Upon explicitly calculating the function $B$ and analytically continuing it in $s$ to the region 
$(m_{1}+m_{2})^2<s<\Lambda $, we arrive at an ultimate expression (5) for $B(s)$.

By substituting the calculated function $B(s)$ into (1), we find the roots $s_{0}$ of Eq (1). 
Equation (1) can be solved numerically, but it is more convenient to use a graphical method to find all 
roots $s_{0}$ in the complex-plane region under investigation. The imaginary and the real part of the 
expression $1-B({\mathrm Re}s,{\mathrm Im}s)$ correspond individually to a certain three-dimensional 
surface above the $s$ plane. A section of each such surface by the $s$ plane gives a certain set of planar 
curves in the z=0 plane. The intersection of these sets of planar curves gives the set of roots $s_{0}$. 
Upon plotting the imaginary and the real part calculated by a computer individually, one finds their intersection 
and visually seeks the roots in the complex-plane region under consideration. The accuracy of the search may be 
improved by enlarging the scale of the graph. Considering that $s_{0}=M^2-iM\Gamma$, we can obtain the mass $M$ 
and the decay width $\Gamma$ as:

$$M^2={\mathrm Re}s_{0},\qquad M\Gamma=-{\mathrm Im}s_{0}. \eqno(6)$$ \\

The calculations show that there is only one root $s_{0}$ in the complex-plane region of interest; 
at a given value of $\Lambda$ the mass and the width of the state corresponding to this root depend only 
on $g$ that is a dimensionless constant from the expression for $G(s)$.

Four parameters were used for the calculation of the $q\bar c$-meson masses: $g_{qc},\; \Lambda_{qc}, \; \Delta_{q}, \; \Delta_{c},$ 
where $g_{qc}$ determines  the constant  $G(s)$, $\Lambda_{qc}$ $-$ a parameter giving the cutoff;
$\Delta_{q}, \; \Delta_{c}$ $-$  $P$-wave mass shifts of the $q$- and $c$-quarks. The effective parameters
$\Delta_{q}$ and $\Delta_{c}$ of quark mass shifts $m_{\mbox {\small{eff}}}  = m + \Delta$ [24] lead to
the scale change of quark potential and allow to obtain bound states in $P$-wave.\\ 
The values of the light quark mass ($m_{u,d}=$0.385 GeV, $m_{s}=$0.510 GeV)  and the heavy quark masses ($m_{c}=$1.645 GeV, 
$m_{b}=$4.94 GeV) were taken from [25].  Four parameters mentioned above were determined by using the experimental 
data on the masses of $D,\ D^{*},\ D_{1}, \ D_{2}^{*}$--mesons. For the computation of $c\bar c$-mesons two more parameters 
were used: $g_{c},\ \Lambda_{c},$  where $g_{c}$ determines $G(s)$, $\Lambda_{c}$ $-$ a parameter giving the cutoff. 
These two parameters were determined by using the experimental data on masses of the $\eta_{c},\ J/\psi$-mesons.
Analogous 
two parameters were used for the $q\bar b$--mesons: $g_{qb},\ \Lambda_{qb},$ where $g_{qb}$ determines  $G(s)$,
$\Lambda_{qb}$ $-$ a parameter giving the cutoff. The parameters were determined by the experimental values of the $B,\; B^{*}$
-meson masses. \\ 
For the $c\bar b$-mesons one parameter -- $g_{cb}$ was used, which is determined by the experimental value of the $B_{c}$-meson 
mass [26].\\
$\Lambda_{cb}$ is derived from $\Lambda_{c}$ and $\Lambda_{b}$: $$\Lambda_{cb}=\dfrac {1} {4} 
(\sqrt{\Lambda_{c}}+\sqrt{\Lambda_{b}})^2.$$
Three parameters were used for the $b\bar b$-mesons: $g_{b}, \ \Lambda_{b},\ \Delta_{b},$\,  where $g_{b}$ determines $G(s)$,
\ $\Lambda_{b}$ $-$ a parameter giving the cutoff. $\Delta_{b}$ determines the $p$-wave mass shift of the $b$-quark.  These parameters 
are fitted on the experimental values of the $\eta_{b}(1S),\ \Upsilon(1S), \ \chi_{b2}(1P)$-meson masses. The results of the computation 
are shown in Tables 1,2. It is seen from Table 1 that for  $D_{sJ}^{*}(2317)$
and $D_{sJ}(2460)$ states the calculation results on masses are 100 MeV lower than experimental values. It seems necessary
to take into account not only $\bar cs$, but tetraquark $\bar csn\bar n$ states also. Such consideration makes
possible to increase the masses of the $D_{sJ}^{*}$ and $D_{sJ}$ states.\\ \\ \\

{\bf 3. Radiative decays of heavy mesons}\\

Calculations of radiative decay widths are carried out with the masses, which are taken from the
experimental tables [26].
$$V\to \gamma P$$
The decay width $V\to \gamma P$: ($V$ corresponds to the notation of a vector meson, $P$ - pseudoscalar meson)\\
{\multlinegap=0pt \begin{multline} 
 \qquad \qquad \;
\Gamma_{V\to \gamma P}=\dfrac {\alpha} {24} \left(1-\left(\dfrac {M_{P}} {M_{V}} \right)^{2} \right) 
\dfrac {1} {M_{V}} \dfrac {|F_{V\to \gamma P}(0)|^{2}} {|F_{V}(0)||F_{P}(0)|}, 
\qquad  \qquad \quad 
\tag{7} \end{multline} }\\
where \; $\alpha=e^{2}/4\pi=1/137, \, M_{V}$ is the vector meson mass, \; $M_{P}$ is the pseudoscalar meson mass.
We are using the dispersion integration method [27]  \\
{\multlinegap=0pt \begin{multline}
F_{V\to \gamma P}(q^{2})=\int\limits_{(m_{1}+m_{2})^{2}}^\infty \dfrac {ds} {\pi} \int
\limits_{(m_{1}+m_{2})^{2}}^\infty \dfrac {ds'} {\pi} \dfrac {disc_{s}disc_{s'}F_{V\to 
\gamma P}(s, s', q^2)} {(s-M_{V}^2)(s'-M_{P}^2)},   \quad \; \, 
\tag{8} \end{multline} }\\
In the center-of-mass frame [28,29]: \\
{\multlinegap=0pt \begin{multline}
F_{V\to \gamma P}(q^{2}) \sim \int\limits_{0}^{k_{max}} {dk} \int\limits_{-1}^{1} {dy} \cdot k^2 \cdot 
\Bigl(D_{1}(s,s_{1}',q^2)S_{V\to \gamma P}(s,s_{1}',q^{2}) \cdot e_{1}f_{1}(q^2) \times 
\\  \shoveleft{   
\qquad \qquad \qquad \qquad \qquad \qquad  \quad
\times \dfrac {G_{P}G_{V}} {(s-M_{V}^2)(s_{1}'-M_{P}^2)} + (1 \rightleftharpoons 2) \Bigr)},  
\quad    
\tag{9} \end{multline}}\\
where $S_{V\to \gamma P}(s,s',q^{2})$  is a spin factor, $G_{P}$ and $G_{V}$  are the vertices, which practically do not
depend on energy, 
$e_{1}$ is the charge of quark 1, $f_{1}$ is its form factor, \\[10pt]
$k_{max}=\sqrt{\dfrac {\Lambda} {4} + \dfrac {(m_{1}^2-m_{2}^2)^2} {4\Lambda} -
\dfrac {1} {2} (m_{1}^2+m_{2}^2) }, \qquad \Lambda=\Lambda_{qQ}\dfrac {(m_{1}+m_{2})^2} {4},$ \\[10pt]
where $k_{max}$ is the quark momentum cutoff,\\
$s=(\sqrt{m_{1}^2+k^2}+\sqrt{m_{2}^2+k^2})^2$ , \quad
$s_{1}'=s-d_{1}q^2-\dfrac{|ky|} {c_{1}}\,\sqrt{q^2s(q^2-4c_{1})}$ ,\\
$d_{1}=\dfrac{1} {c_{1}} (\sqrt{(m_{1}^2+k^2)(m_{2}^2+k^2)}+k^2y^2)\, , 
\quad c_{1}=m_{1}^2+k^2(1-y^2) \,$ , \\
$D_{1} = \dfrac {s} {2c_{1} \sqrt{m_{2}^2+k^2}} \Biggl(1+\dfrac{ky\sqrt{-q^2}} 
{\sqrt{(m_{1}^2+k^2)(4c_{1}-q^2)}}\Biggr).$ \\
 
{\multlinegap=0pt \begin{multline}
F_{V}(q^{2}) \sim \int\limits_{0}^{k_{max}} {dk} \int\limits_{-1}^{1} {dy} \cdot k^2 \cdot 
\Bigl(D_{1}(s,s_{1}',q^2)S_{V}^{tr}(s,s_{1}',q^{2}) \cdot e_{1}f_{1}(q^2) \times  
\qquad \qquad \qquad  \qquad \qquad \\ \shoveleft{
 \qquad \qquad \qquad  \qquad \qquad \qquad \qquad \qquad 
\times \dfrac {G_{V}^2} {(s-M_{V}^2)(s_{1}'-M_{V}^2)} + (1 \rightleftharpoons 2)
\Bigr),  }  \qquad  \qquad  \raisetag{26pt} \tag{10} \end{multline} } \\[-30pt]
{\multlinegap=0pt \begin{multline}
F_{P}(q^{2}) \sim \int\limits_{0}^{k_{max}} {dk} \int\limits_{-1}^{1} {dy} \cdot k^2 \cdot 
\Bigl(D_{1}(s,s_{1}',q^2)S_{P}^{tr} (s,s_{1}',q^{2}) \cdot e_{1}f_{1}(q^2)  \times  
\qquad \qquad \qquad \qquad \qquad \\  \shoveleft{
\qquad \qquad \qquad \qquad \qquad \qquad \qquad \qquad
\times \dfrac {G_{P}^2} {(s-M_{P}^2)(s_{1}'-M_{P}^2)} + (1 \rightleftharpoons 2)
\Bigr),}   \qquad  \qquad    \raisetag{26pt}  \tag{11} \end{multline}}
\\
The expressions for the spin factors [28, 29]:\\
{\multlinegap=0pt \begin{multline}
S_{V \to \gamma P} (s,s',q^2)=8 \Biggl\{m_{1} \; \dfrac {  (s'-s)\dfrac{s'-s-q^2} {2} -q^2(s+m_{2}^2-m_{1}^2)}
{(s'-s)^2-2q^2(s'+s)+q^4} \; -  \qquad \qquad \qquad \qquad \qquad \qquad \qquad \qquad
\\[12pt] \shoveleft {\qquad \qquad \qquad \qquad \qquad \qquad \qquad
 - \; m_{2}q^2 \; \dfrac { \dfrac{s'+s-q^2} {2} + m_{1}^2-m_{2}^2 } 
{(s'-s)^2-2q^2(s'+s)+q^4} \Biggr\},}
\\[30pt]
\shoveleft{
S_{P}^{tr}(s,s',q^2)=-q^2 \Biggl\{ \dfrac {(s'+s-q^2+2(m_{1}^2-m_{2}^2))(s+s'-2(m_{1}-m_{2}^2))}
{(s'-s)^2-2q^2(s'+s)+q^4}+ \qquad \qquad \qquad \qquad \qquad} \\[12pt] \shoveleft{ 
\qquad \qquad \qquad \qquad \qquad \qquad \qquad \qquad   + q^2 \dfrac {s'+s-
\dfrac {(s'-s)^2} {q^2} +2(m_{2}^2-m_{1}^2) } {(s'-s)^2-2q^2(s'+s)+q^4} \Biggr\}  ,}  \qquad 
\raisetag{125pt} \tag{12} \end{multline}}\\
{\multlinegap=0pt \begin{multline}
S_{V}^{tr}(s,s',q^2)=-\dfrac{2} {3} q^2 \Biggl(1+\frac{s+s'-q^2+2(m_{1}^2-m_{2}^2)} {4q^2s-(s'-s-q^2)^2} \Bigl[
s+s'-q^2+4m_{1}m_{2} + \frac{1} {2} \frac {m_{2}-m_{1}} {ss'} \times \\ \times \Bigl\{(s+s'-q^2)
\Bigl((m_{2}+m_{1})(s+s')-\frac{1} {2}(m_{2}-m_{1})(s+s'-q^2-2(m_{2}+m_{1})^2 \Bigr)-\\ -2(m_{2}+m_{1})
(2ss'+(m_{2}^2-m_{1}^2)(s+s'))\Bigr\} \Bigr]+\\ \qquad \qquad
+\frac {1} {2} \frac {m_{2}-m_{1}} {ss'} \Bigl[m_{1}(s+s')+m_{2}q^2+\frac {1} {2}(m_{2}+m_{1})(s+s'-q^2)
\Bigr] \Biggr) \qquad \qquad \qquad \quad
\raisetag{60pt} \tag{12} \end{multline}}
The numerical results:
$$B^{*}(5325) \to \gamma B(5279)$$
$k_{max}=1.163$ GeV, $\Lambda=39.70$ GeV$^2$\\
$\Gamma_{B^{*}(5325) \to \gamma B(5279)} \approx 0.8$ keV \\
$$A\to \gamma V$$
The decay width $A \to \gamma V$: ($V$ corresponds to the notation of a vector meson, $A-$ the axial meson)
{\multlinegap=0pt \begin{multline}
\quad  \Gamma_{A \to \gamma V}=\dfrac {\alpha} {12} \left(1-\left(\dfrac{M_{V}} {M_{A}} \right)^2 \right) 
\dfrac {1} {M_{A}} \dfrac {M_{A}^4+6M_{A}^2M_{V}^2+M_{V}^4} {2M_{V}^2} \dfrac {|F_{A \to \gamma V}(0)|^2} 
{|F_{V}(0)||F_{A}(0)|}, \qquad \quad   
\raisetag{30pt} \tag{13} \end{multline}} \\
where $M_{V}$ is the vector meson mass, $M_{A}$ is the axial meson mass .
{\multlinegap=0pt \begin{multline}
\qquad  
F_{A\to \gamma V}(q^{2})=\int\limits_{(m_{1}+m_{2})^{2}}^\infty \dfrac {ds} {\pi} \int\limits_{(m_{1}+m_{2})^{2}}^\infty 
\dfrac {ds'} {\pi} \dfrac {disc_{s}disc_{s'}F_{A\to \gamma V}(s, s', q^2)} {(s-M_{A}^2)(s'-M_{V}^2)}, \qquad \qquad \quad  
\raisetag{40pt} \tag{14} \end{multline}}\\[-10pt]
In the center-of-mass frame [28,29]:\\[-20pt]
{\multlinegap=0pt \begin{multline}
F_{A\to \gamma V}(q^{2}) \sim \int\limits_{0}^{k_{max}} {dk} \int\limits_{-1}^{1} {dy} \cdot k^2 \cdot \Bigl(D_{1}(s,s_{1}',q^2)S_{A\to 
\gamma V}(s,s_{1}',q^{2}) \cdot e_{1}f_{1}(q^2) \times \qquad \qquad \qquad \qquad \\[-15pt]  \shoveleft{
\qquad \qquad \qquad \qquad \qquad \qquad  \qquad  \quad \qquad \times
\dfrac {G_{A}G_{V}} {(s-M_{A}^2)(s_{1}'-M_{V}^2)} + (1 \rightleftharpoons 2) \Bigr), } \qquad  \quad \;   
\raisetag{26pt} \tag{15} \end{multline}}\\
where $S_{A\to \gamma V}(s,s',q^{2})$ is the spin factor, $G_{A}$ and $G_{V}$ are the vertices, which practically do not
depend on energy, 
$e_{1}$ is the charge of quark 1, $f_{1}$ is its form factor, 
{\multlinegap=0pt \begin{multline}
F_{A}(q^{2}) \sim \int\limits_{0}^{k_{max}} {dk} \int\limits_{-1}^{1} {dy} \cdot k^2 \cdot \Bigl(D_{1}(s,s_{1}',q^2)S_{A}^{tr}
(s,s_{1}',q^{2}) \cdot e_{1}f_{1}(q^2) \times  \qquad \qquad \qquad  \qquad \qquad
 \\ \qquad \qquad \qquad  \qquad \qquad \qquad \qquad \qquad \qquad \qquad \times 
\dfrac {G_{A}^2} {(s-M_{A}^2)(s_{1}'-M_{A}^2)} + (1 \rightleftharpoons 2)
\Bigr),    \qquad \quad 
\raisetag{26pt} \tag{16} \end{multline}}\\
The expressions for the spin factors [28, 29]:
{\multlinegap=0pt \begin{multline}
S_{A \to \gamma V} (s,s',q^2)=\dfrac{1} {z_{11}(s,s',q^2)} \cdot 4i \cdot \sqrt{\dfrac{2} {3s}} \Biggl\{m_{1}^2 \Biggl( \biggl(
\frac {s'-s} {2} \biggr)^2 - m_{2}^2q^2 \Biggr) + \frac{q^2} {2} \Biggl(m_{2}^2q^2+
\qquad \quad  \\[20pt] \shoveleft{+2\biggl(\frac{s'} {2} - \frac{m_{1}^2+m_{2}^2}
{2} \biggr) \biggl(\frac{s} {2} - \frac{m_{1}^2+m_{2}^2} {2} \biggr) \Biggr) \Biggr\} \Biggl\{\frac{(m_{1}+m_{2})(s-s'+q^2)
(s+s'-q^2+2m_{1}^2-2m_{2}^2)} {4q^2s-(s-s'+q^2)^2}+}\\[20pt] \shoveleft{
+\frac{m_{1}-m_{2}} {s'} \biggl(\frac{s+s'-q^2} {2} - (m_{1}+m_{2})^2\biggr) \Biggr\} ,} \qquad \qquad \qquad 
\qquad \qquad \qquad \qquad \qquad \qquad \qquad  
\raisetag{26pt} \tag{17} \end{multline}}
where \\
$$z_{11} (s,s',q^2)=\dfrac{-(s^2+6ss'+s'^2)+q^2(2s+2s'-q^2)} {2s'} ,$$ 
{\multlinegap=0pt \begin{multline}
S_{A}^{tr} (s,s',q^2) =\dfrac{\dfrac{4} {3} q^2s'} {4q^2s-(s-s'+q^2)^2}
\Biggl\{m_{1}^2\Biggl(\biggl(\dfrac{s'-s} {2}\biggr)^2-m_{2}^2q^2\Biggr)+
\qquad \qquad \qquad \qquad \qquad\\ \qquad \qquad \qquad \qquad \qquad \quad + 
\dfrac{q^2} {2} \Biggl(m_{2}^2\dfrac{q^2}{2}+2\biggl(\dfrac{s'}{2}-\dfrac{m_{1}^2+m_{2}^2} {2} \biggr)
\biggl(\dfrac{s}{2}-\dfrac{m_{1}^2+m_{2}^2} {2} \biggr)\Biggr)\Biggr\}  \qquad \qquad 
\raisetag{26pt} \tag{18} \end{multline}}\\
The numerical results:
$$D_{sJ}(2460) \to \gamma D_{s}^{*}(2112)$$
$k_{max}=1.139$ GeV, $\Lambda=12.97$ GeV$^2$ ($m_{1}=2030.5$ MeV, $m_{2}=565$ MeV)\\
$\Gamma_{D_{sJ}(2460) \to \gamma D_{s}^{*}(2112)} \approx 5.73$ keV.

$$S \to \gamma V$$
The decay width $S \to \gamma V$: ($V$ corresponds to the notation of a vector meson, $S-$ the scalar meson)  
{\multlinegap=0pt \begin{multline}
\qquad  \qquad  \qquad  
\Gamma_{S \to \gamma V}=\dfrac{\alpha} {2} \left(1-\left(\dfrac{M_{V}} {M_{S}} \right)^2 \right) 
\dfrac {1} {M_{S}} \dfrac {|F_{S \to \gamma V}(0)|^2} {|F_{V}(0)||F_{S}(0)|}, 
\qquad  \qquad  \qquad  \qquad  \quad \; \; 
\raisetag{26pt} \tag{19} \end{multline}}
where $M_{V}$ is the vector meson mass, $M_{S}$ is the scalar meson mass, which is above threshold, i.e. $M_{S}>(m_{1}+m_{2})^2$.
The Lehmann representation [30] is used here 
{\multlinegap=0pt \begin{multline}
\qquad    
F_{S \to \gamma V}(q^2)=\int\limits_{(m_{1}+m_{2})^2}^{\infty} {\dfrac{ds} {\pi}} \int
\limits_{(m_{1}+m_{2})^2}^{\infty} {\dfrac {ds'} {\pi} \dfrac {1} {D_{R}(s)} 
\dfrac{disc_{s}disc_{s'}F_{S \to \gamma V}(s,s',q^2)} {s'-M_{V}^2}}.  
\qquad \qquad 
\raisetag{26pt} \tag{20} \end{multline}}
The inverse propagator
{\multlinegap=0pt \begin{multline} 
\qquad \qquad \qquad
D_{R}(m^2)=m_{R}^2-m^2+Re\bigl(\Pi_{R}(m_{R}^2)\bigr)-\Pi_{R}(m^2),  
\qquad \qquad \qquad \qquad \qquad \quad 
\raisetag{26pt} \tag{21} \end{multline}}
where $Re\bigl(\Pi_{R}(m_{R}^2)\bigr)-\Pi_{R}(m^2)$ takes into account the finite width corrections of the 
scalar  $R$ resonance, which take into account the contribution of the two-particle virtual intermediate $ab$ states to
self-energy of the $R$ resonance,
{\multlinegap=0pt \begin{multline}
\qquad \qquad \qquad \qquad \qquad \qquad \qquad 
\Pi_{R}(m^2)=\sum\limits_{ab}^{}\Pi_{R}^{ab}(m^2). \qquad \qquad \qquad 
\qquad \qquad \qquad \; 
\raisetag{26pt} \tag{22} \end{multline}}\\
In real axis of $m^2$
{\multlinegap=0pt \begin{multline}
\qquad Im\bigl(\Pi_{R}(m^2)\bigr)=m\Gamma_{R}(m)=m\sum\limits_{ab} \Gamma(R \to ab, m) \theta(m-m_{a}-m_{b}), 
\qquad \qquad \quad \; \;
\raisetag{26pt} \tag{23} \end{multline}} 
where
{\multlinegap=0pt \begin{multline}
\qquad \qquad \qquad \qquad \qquad \quad 
\Gamma(R \to ab, m) = \dfrac{g_{R_{ab}}^2} {16\pi m} \rho_{ab}(m^2), 
\qquad \qquad \qquad \qquad \qquad \; 
\raisetag{26pt} \tag{24} \end{multline}} 
is the width of the $R \to ab$ decay, $m=m_{ab}$ is the invariant mass of the $ab$ state, 
$g_{R_{ab}}$ is the coupling constant of the $R$ resonance with the two-particle $ab$ state, and 
{\multlinegap=0pt \begin{multline}
\qquad \qquad \quad \; \;
\rho_{ab}(m^2)=\sqrt{\biggl(1-\dfrac{m_{+}^2} {m^2} \biggr) \biggl(1- \dfrac {m_{-}^2} {m^2}\biggr)}, 
\quad m_{\pm}=m_{a} \pm m_{b}. \qquad \qquad \qquad 
\raisetag{26pt} \tag{25} \end{multline}}  

Propagators under discussion satisfy the Lehmann representation 
{\multlinegap=0pt \begin{multline}
\dfrac{1} {D_{R}(m^2)}=\dfrac{1} {\pi} \int \limits_{(m_{a}+m_{b})^2}^{\infty} {\dfrac{Im\bigl(
\frac{1} {D_{R}(\bar m^2)}\bigr)} {\bar m^2-m^2-i\epsilon} d\bar m^2}= \\[-10pt] 
\qquad \qquad \qquad \qquad \qquad \qquad    =\dfrac{1} {\pi} \int 
\limits_{(m_{a}+m_{b})^2}^{\infty} {\dfrac{\bar m\Gamma_{R}(\bar m)}
{|D_{R}(\bar m^2)|^2 (\bar m^2-m^2-i\epsilon)} d\bar m^2 } \qquad \qquad
\raisetag{38pt} \tag{26} \end{multline}} 
in the wide domain of coupling constants of the scalar $R$ resonance with the two-particle $ab$ states.

In the one-channel case when the decay threshold is lower than the $R$ resonance one has: 
{\multlinegap=0pt \begin{multline}
\Pi_{R}(m^2)=\Pi_{R}^{ab}(m^2) =\dfrac{g_{R_{ab}}^2} {16\pi^2} \Biggl\{\dfrac{(m^2-m_{+}^2)} {m^2} 
\dfrac{m_{-}} {m_{+}} \ln {\dfrac{m_{a}} {m_{b}}}+ \qquad \qquad \qquad  \qquad \qquad \qquad\\
\qquad \qquad \qquad \qquad  \qquad \qquad   +\rho_{ab}(m^2) \Biggl[i\pi+\ln{\dfrac{
\sqrt{m^2-m_{-}^2}-\sqrt{m^2-m_{+}^2}} {\sqrt{m^2-m_{-}^2}+\sqrt{m^2-m_{+}^2}}}\Biggr]\Biggr\} 
\qquad \quad
 \raisetag{30pt} \tag{27} \end{multline}}
at $m\ge m_{+}=m_{a}+m_{b},$ \; $(m_{a}\ge m_{b})$. \\

In the  center-of-mass frame:
{\multlinegap=0pt \begin{multline}
F_{S \to \gamma V}(q^2) \sim \int_{0}^{k_{max}} {dk} \int_{-1}^{1}{dy} \cdot k^2 \cdot \dfrac {1} {D_{R}(s)}
 \times \qquad \qquad \qquad \qquad \qquad \qquad \qquad \qquad \qquad \qquad \qquad \qquad \\
\qquad \qquad \qquad \times  \Biggl(
D_{1}(s,s_{1}',q^2)S_{S \to \gamma V}(s,s_{1}',q^2) \cdot e_{1}f_{1}(q^2)\cdot 
\dfrac {G_{S}G_{V}} {s_{1}'-M_{V}^2} +(1\rightleftharpoons2) \Biggr),  \qquad \quad  
 \raisetag{28pt} \tag{28} \end{multline}} \\[-40pt]
{\multlinegap=0pt \begin{multline}
F_{S}(q^2) \sim
\int_{0}^{k_{max}} {dk} \int_{-1}^{1}{dy} \cdot k^2 \cdot \dfrac {1} {D_{R}(s)} \times   
\qquad \qquad \qquad \qquad \qquad \qquad \qquad \qquad \qquad \qquad \qquad \qquad \\ 
\qquad \qquad \qquad \; \; \times \Biggl(
D_{1}(s,s_{1}',q^2)S_{S}^{tr}(s,s_{1}',q^2) \cdot e_{1}f_{1}(q^2)\cdot \dfrac {G_{S}^2} {s_{1}'-M_{S}^2}
+(1\rightleftharpoons2) 
\Biggr),  \qquad \qquad 
 \raisetag{28pt} \tag{29} \end{multline}}
The expressions for the spin factors:
{\multlinegap=0pt \begin{multline}
S_{S \to \gamma V}(s,s',q^2)=\dfrac {1} {s'}(m_{2}-m_{1})(s+m_{1}^2-m_{2}^2-q^2)(s'-(m_{1}+m_{2})^2)\; -  
\qquad \qquad \qquad\\[5pt]
-m_{1} \left(4m_{1}m_{2}-\dfrac{5} {2}m_{1}^2-\dfrac{3} {2} m_{2}^2+2s' \right)
+\dfrac{1} {s'} \dfrac{2s'q^2+(s'-s+q^2)(s+m_{1}^2-m_{2}^2-q^2)} {4s'q^2-(s'-s+q^2)^2} \times \\[5pt]
\shoveleft{\times \{2s'(m_{1}(s'-s)-
m_{2}q^2)+(m_{2}-m_{1})(s'-s+q^2)(s'-(m_{1}+m_{2})^2)\},}\\
 \qquad \qquad \qquad \qquad \qquad \qquad \qquad \qquad \qquad \qquad \qquad \qquad \qquad \qquad 
 \qquad \qquad \qquad \qquad   \\
S_{S}^{tr}(s,s',q^2)=4\dfrac{q^2(ss'-m_{2}(m_{1}+m_{2})(s+s'-q^2)-(m_{1}+m_{2})^2(m_{1}^2-m_{2}^2))} 
{4q^2s-(s-s'+q^2)^2} \qquad  
 \raisetag{75pt} \tag{30} \end{multline}}
The numerical results:
$$D_{sJ}^{*}(2317)\to \gamma D_{s}^{*}(2112)$$
$k_{max}=1.135$ GeV, $\Lambda=8.94$ GeV$^2$, $g_{R \to ab} \approx 1.5$ GeV$^2$.\\
$\Gamma_{D_{sJ}^{*}(2317)\to \gamma D_{s}^{*}(2112)} \approx 2.2$ keV. \\[10pt]
For comparison the calculation results on these radiative decays in other models are 
presented in table 5. \\[-20pt]

{\bf 4. Conclusion}\\

    The relativistic quark model for the study of the heavy meson spectroscopy is constructed in the framework of the dispersion approach. 
The N/D method of heavy quark effective theory was used for the computation of the mass spectrum of the charmed and beauty mesons: the multiplets $J^{PC}=0^{-+}, 1^{--} (S$-wave) 
and $J^{PC}=0^{++}, 1^{+-}, 1^{++}, 2^{++}$ ($P$-wave). The $D_{sJ}^{*}(2317)$ and $D_{sJ}(2460)$ states attract a particular attention. 
The masses of these states are obtained to be 100 MeV smaller as compared to the experimental data (Table 1).
It seems  not enough to account the contribution of the  $\bar cs$ configuration only, but it is
necessary to take into account a composite of tetraquark state [31] to obtain the masses close to the experimental data.
The radiative decay widths have been calculated for these mesons using the experimental data on their masses. The radiative decay widths of these states are in  good agreement with the
results of other quark models (Table 5) [32, 33]. Table 5 contains also the  calculation results  in the framework of the Model of QCD 
Sum Rules [34] and of the Vector Dominance Model [35]. In the present work we managed to describe 
all $S$- and $P$-wave multiplets of the charmed and beauty mesons using a small number of parameters. The proposed method allows
to satisfy in the case of $b$-mesons the qualitative ratio for gluon coupling constants: $g_{qb}<g_{b}$. The use of the
dispersion N/D method in heavy quark effective theory allows to consider the spectroscopy of exotic heavy mesons and
baryons taking into account the threshold singularity. \\ 
\\ 

    The authors would like to thank T. Barnes, Fl. Stancu and S. V. Chekanov for useful discussions. This research was supported by Russian 
Ministry of Education (Grant 2.1.1.68.26).

\newpage
{\bf References}\\ 

1.	J. M. Richard, Phys. Lett. B {\bf 139}, 408 (1984).

2.	A. A. Martin,  Phys. Lett. B {\bf 100}, 511 (1981).

3.	J. L. Basdevant, and S. Boukraa,  Z. Phys. C {\bf 30}, 103 (1986).
 
4.	H. W. Crater, and P. V. Alstine, Phys. Rev. D {\bf 37}, 1982 (1988).

5.	K. G. Boreskov and A. B. Kaidalov, Yad. Fiz. {\bf 37}, 174 (1983)

6.	C. Quigg, J. L. Rosner, Phys. Rep. C {\bf 56}, 167 (1979).

7.	A. A. Bykov, I. M. Dremin and A. V. Leonidov, Usp. Fiz. Nauk {\bf 143},

 3 (1984)

8.	M. A. Shifman, Usp. Fiz. Nauk {\bf151}, 193 (1987)

9.	M. B. Voloshin, Yad. Fiz. {\bf 29}, 1368 (1979)

10.	Babar Collaboration, B. Aubert {\em et al.}, Phys. Rev. Lett. {\bf 90}, 242001 

(2003).

11.	Belle Collaboration, P. Krokovny {\em et al.}, Phys. Rev. Lett. {\bf 91}, 262002 

(2003).

12.	Belle Collaboration, Y. Mikami {\em et al.}, Phys. Rev. Lett. {\bf 92}, 012002 

(2004).

13.	Belle Collaboration, A. Drutskoy {\em et al.}, Phys. Rev. Lett. {\bf 94}, 061802 

(2005).

14.	Babar Collaboration, G. Calderini {\em et al.}, hep-ex/0405081.

15.	P. Colangelo {\em et al.}, hep-ph/0512083.

16.	W. Wei, P-Z. Huang and S.-L. Zhu, hep-ph/0510039.

17.	T. Barnes, F. E. Close, and H. J. Lipkin, Phys. Rev. D {\bf 68}, 054006 

(2003).

18.	H.-J. Cheng and W.-S. Hou, Phys. Lett. B {\bf 566}, 193 (2003).

19.	K. Terasaki, Phys. Rev. D {\bf 68}, 011501 (2003).

20.	A. P. Szszepaniak, Phys. Lett. B {\bf 567}, 23 (2003).

21.	V. V. Anisovich, S. M. Gerasyuta, I. V. Keltuyala, Yad. Fiz. {\bf 38}, 200 

(1983)
 
22.	V. V. Anisovich, S. M. Gerasyuta, Yad. Fiz. {\bf 44}, 174 (1986)

23.	V. V. Anisovich, S. M. Gerasyuta and A. V. Sarantsev,  Int. J. Mod. 

Phys. A {\bf 6}, 625 (1991).

24.	S. M. Gerasyuta, I. V. Keltuyala, Yad. Fiz. {\bf 54}, 793 (1991)

25.	S. M. Gerasyuta, A. V. Sarantsev, Yad. Fiz. {\bf 52}, 1483 (1990)

26.	W.-M. Yao {\em et al.}(Particle Data Group), J. Phys. G {\bf 33}, 1 (2006).

27. V. V. Anisovich, A. V. Sarantsev, Yad. Fiz. {\bf 45}, 1479 (1987)

28.	A. V. Anisovich, V. V. Anisovich and V. A. Nikonov,  hep-ph/0108186.
  
29.	A. V. Anisovich, V. V. Anisovich, V. N. Markov, M. A. Matveev, N. 

A. Nikonov, and A. V.  Sarantsev, hep-ph/0509042.

30.	N. N. Achasov and A. V. Kiselev, hep-ph/0405128.

31. S. M. Gerasyuta, V. I. Kochkin, Yad. Fiz. {\bf 61}, 1504 (1998)

32.	S. Godfrey, Phys. Lett. B {\bf 568}, 254 (2003).

33.	W. A. Bardeen, E. J. Eichten and C. T. Hill, Phys. Rev. D {\bf 68}, 

054024 (2003).

34.	P. Colangelo, F. De Fazio and A. Ozpineci, Phys. Rev. D {\bf 72}, 074004 

(2005).

35.	P. Colangelo, F. De Fazio, and R. Ferrandes, Mod. Phys. Lett. A 

{\bf 19}, 2083 (2004).

 \newpage
\noindent Table 1. The lowest states of the charmonium and states with open charm masses. \\[8pt]  
\begin{tabular} {|c|c|c|c|}
\hline
\qquad & $m(0^{-+})$, GeV  & $m(1^{--})$, GeV & $m(0^{++})$, GeV    \\
\hline
 $u\bar c \quad d\bar c$  & $D \quad \quad \ \ $ 1.867 (1.867) & $D^{*} \quad$ 2.010 (2.010) & $ \ D_{0}^{*}  \qquad \quad  $ 2.103 (2.352$\pm$0.05)  \\
 \hline
 $s\bar c$ & $D_{s}\quad \quad  \ $ 1.916 (1.969) & $D_{s}^{*} \quad$ 2.105 (2.112) & $D_{sJ}^{*}  \quad \quad \ \ $ 2.223 (2.317) \qquad \qquad   \\
 \hline
 $c\bar c$ & $\eta_{c}(1S)\quad$ 2.980 (2.980) & $J/\psi \ \ $ 3.097 (3.097) & $\chi_{c0}(1P) \quad $ 3.393 (3.415) \qquad \qquad  \\
\hline
\end{tabular}
\begin{tabular} {|c|c|c|c|}
\hline
\qquad  & $m(1^{+-})$, GeV & $m(1^{++})$, GeV & $m(2^{++})$, GeV   \\
\hline
 $u\bar c \quad d\bar c$  & $D_{1} \quad$ 2.302 (2.420) & $D_{1}(2430) \quad$ 2.430 (2.430) & $D_{2}^{*}\qquad \quad \ $ 2.460 (2.460) \\
 \hline
 $s\bar c$  &$D_{sJ}\ \ $ 2.350 (2.460) & $ D_{s1} \qquad \quad \ \ $ 2.514 (2.536) & $D_{s2}\qquad \quad$ 2.559 (2.573) \\
 \hline
 $c\bar c$  &\ \ \  3.726 (-) & $\ \chi_{c1}(1P)\quad \ \ $ 3.824 (3.511) & $\chi_{c2}(1P) \quad $ 3.863 (3.556)   \\
\hline
\end{tabular}\\ \\ \\
The model parameters:\qquad \qquad $g_{qc}=1.26,\quad \Lambda_{qc}=7.90, \quad \Delta_{q}=0.035$ GeV$, \\ \Delta_{c}=0.404$ GeV, $g_{c}=2.92,\quad 
\Lambda_{c}=5.49$.

The experimental data are given in parentheses [26].
\newpage

\noindent Table 2. The lowest states of the bottomonium and states with open bottom masses. \\[8pt] 
\begin{tabular} {|c|c|c|c|}
\hline
\qquad & $m(0^{-+})$, GeV  & $m(1^{--})$, GeV & $m(0^{++})$, GeV    \\
\hline
 $u\bar b \quad d\bar b$  & $B \quad \ \ \ $ 5.279 (5.279)\qquad \qquad \  & $B^{*}\quad \ $ 5.325 (5.325) &\quad  \ \ 5.508 (-)  \\
 \hline
 $s\bar b$ & $B_{s} \quad \ \ $ 5.339 (5.370)\qquad \qquad \  & $B_{s}^{*} \quad \ $ 5.419 (5.417) &\quad \ \  5.629 (-)  \\
 \hline
 $c\bar b$  & $\ B_{c}\quad  \ \ $ 6.400 (6.4$\pm$0.39$\pm$0.13) & \ \ \  6.477 (-) &\quad \ \  6.818 (-)  \\
 \hline
 $b\bar b$ & $\eta_{b}(1S) \  $ 9.330 (9.300$\pm$0.040) \quad \    & $\Upsilon(1S) \  $ 9.460 (9.460) & $\chi_{b0}(1P)$ 10.139 (9.859)   \\
\hline
\end{tabular}
\begin{tabular} {|c|c|c|c|}
\hline
\qquad  & $m(1^{+-})$, GeV & $m(1^{++})$, GeV & $m(2^{++})$, GeV   \\
\hline
 $u\bar b \quad d\bar b$  &  5.534 (-) &\qquad \ 5.579 (-) & \qquad \ 5.589 (-) \\
 \hline
 $s\bar b$  & 5.595 (-) &\qquad \ 5.664 (-) &\qquad \ 5.685 (-) \\
 \hline
 $c\bar b$   & 6.973 (-) &\qquad \ 7.046 (-) &\qquad \ 7.073 (-)  \\
 \hline
 $b\bar b$  & 9.767 (-) & $\chi_{b1}(1P)\ $ 9.870 (9.893) & $\chi_{b2}(1P)\ $ 9.912 (9.912)   \\
\hline
\end{tabular}\\ \\ \\
The model parameters: \quad $g_{qb}=3.36,\quad \Lambda_{qb}=5.70,\quad g_{cb}=2.92,\quad g_{b}=4.07, \\ \Lambda_{b}=4.9,\quad \Delta_{b}=0.232$ GeV.

The experimental data are given in parentheses [26].

\newpage
Table 3. \quad Coefficients \quad $\alpha,\quad \beta, \quad \delta$.\\[8pt]
\begin{tabular} {|c|c|c|c|}
\hline
$J^{PC}$ & $\alpha$ &  $\beta$  &  $\delta$  \\
\hline 
 $0^{-+}$ & $\dfrac {1} {2}$ & $-\dfrac {1} {2} \dfrac {(m_{1}-m_{2})^2} {(m_{1}+m_{2})^2}$ & 0 \\ 
\hline
 $1^{--}$ & $\dfrac {1} {3}$ & $\dfrac {1} {6}-\dfrac {1} {3} \dfrac {(m_{1}-m_{2})^2} {(m_{1}+m_{2})^2}$ & $-\dfrac {1} {6} (m_{1}-m_{2})^2$ \\
\hline
 $0^{++}$ & $-\dfrac {1} {2}$ & $\dfrac {1} {2}$ & 0  \\
\hline
 $1^{+-}$ &$\dfrac {1} {2}$  & $-\dfrac {1} {2}\dfrac {(m_{1}-m_{2})^2} {(m_{1}+m_{2})^2}$ & 0   \\
\hline
 $1^{++}$ & $\dfrac {1} {2}$ & $-\dfrac {1} {2} \dfrac {(m_{1}-m_{2})^2} {(m_{1}+m_{2})^2}$ & 0  \\
\hline
 $2^{++}$ & $\dfrac {3} {10}$ & $\dfrac {1} {5} \left(1-\dfrac {3} {2} \dfrac {(m_{1}-m_{2})^2} {(m_{1}+m_{2})^2}\right)$ & $-\dfrac {1} {5}(m_{1}-m_{2})^2$   \\
\hline
\end{tabular}\\ \\ \\
\newpage
\noindent Table 4. Functions $G(s)$. \\[8pt]
\begin{tabular} {|c|c|}
\hline
$J^{PC}$ & $G(s)$  \\
\hline
 $0^{-+}$ & $\dfrac {8g} {3} -\dfrac {4g} {3} \dfrac {(m_{1}+m_{2})^2} {s}$ \\
\hline
 $1^{--}$& $\dfrac {4g} {3}$  \\
\hline
 $0^{++}$& $-\dfrac {8g} {3} +\dfrac {4g} {3} \dfrac {(m_{1}-m_{2})^2} {s}$   \\
\hline
 $1^{+-}$ & $\dfrac {8g} {3} -\dfrac {4g} {3} \dfrac {(m_{1}+m_{2})^2} {s}$  \\
\hline
 $1^{++}$ & $\dfrac {4g} {3}$  \\
\hline
 $2^{++}$ & $\dfrac {4g} {3}$  \\
\hline
 \end{tabular}\\ \\ 
\newpage
\noindent Table 5. The radiative decay widths of the charmed mesons (in keV)\\[8pt]
\begin{tabular} {|c|c|c|c|c|c|}
\hline
Decay & LCSR [34] & VMD [35] & QM [32] & QM [33] & [*] \\
\hline
$D_{sJ}^{*}(2317) \to \gamma D_{s}^{*}$ & 4-6 & 0.85 & 1.9 & 1.74 & 2.2 \\
\hline
 $D_{sJ}(2460) \to \gamma D_{s}^{*}$ & 0.6-1.1 & 1.5 & 5.5 & 4.66 & 5.73 \\
 \hline
 \end{tabular}\\ \\
 The symbol [*] denotes the results of this study.

\end{document}